\documentclass[aps,pra,twocolumn,showpacs,superscriptaddress]{revtex4}
\usepackage{amsmath,amssymb,amsfonts,bm,graphicx,latexsym}
\usepackage[hypertex,breaklinks=true,colorlinks=false]{hyperref}
\newcommand{\bra}[1]{\langle #1|}
\newcommand{\ket}[1]{|#1 \rangle}
\newcommand{\bracket}[2]{\langle #1|#2 \rangle}

\newcommand{\ua}{\uparrow}
\newcommand{\da}{\downarrow}

\newcommand{\Scal}{{\cal S}}
\newcommand{\group}{\mathrm{group}}

\newcommand{\rv}{\bm{r}}

\newcommand{\pv}{\bm{p}}

\newcommand{\Kv}{\bm{K}}

\newcommand{\Lv}{\bm{L}}

\newcommand{\Psih}{\hat{\Psi}}

\newcommand{\Psit}{\tilde{\Psi}}

\newcommand{\Pf}{\mathrm{Pf}}

\newcommand{\ev}{\bm{e}}

\newcommand{\PRL}[3]{Phys. Rev. Lett. {\bf #1}, \href{https://doi.org/10.1103/PhysRevLett.#1.#2}{#2} (#3)}
\newcommand{\PRX}[3]{Phys. Rev. X {\bf #1}, \href{https://doi.org/10.1103/PhysRevX.#1.#2}{#2} (#3)}
\newcommand{\PRA}[3]{Phys. Rev. A {\bf #1}, \href{https://doi.org/10.1103/PhysRevA.#1.#2}{#2} (#3)}
\newcommand{\PRB}[3]{Phys. Rev. B {\bf #1}, \href{https://doi.org/10.1103/PhysRevB.#1.#2}{#2} (#3)}
\newcommand{\PRE}[3]{Phys. Rev. E {\bf #1}, \href{https://doi.org/10.1103/PhysRevE.#1.#2}{#2} (#3)}
\newcommand{\PRAR}[3]{Phys. Rev. A {\bf #1}, \href{https://doi.org/10.1103/PhysRevA.#1.#2}{#2} (R) (#3)}
\newcommand{\PRBR}[3]{Phys. Rev. B {\bf #1}, \href{https://doi.org/10.1103/PhysRevB.#1.#2}{#2} (R) (#3)}
\newcommand{\RMP}[3]{Rev. Mod. Phys. {\bf #1}, \href{https://doi.org/10.1103/RevModPhys.#1.#2}{#2} (#3)}

\newcommand{\arXiv}[1]{arXiv:\href{http://arxiv.org/abs/#1}{#1}}

\begin{document}

%%%%%%%%%%%%%%%%%%%%%%%%%%%%%%%%%%%%%%%%%%%%%%%%%
% Paper Information
%%%%%%%%%%%%%%%%%%%%%%%%%%%%%%%%%%%%%%%%%%%%%%%%%
\title{
Quantum Hall phase diagram of two-component Bose gases: \\
Intercomponent entanglement and pseudopotentials
%in a synthetic magnetic field
}
\author{Shunsuke Furukawa}
\affiliation{Department of Physics, University of Tokyo, 7-3-1 Hongo, Bunkyo-ku, Tokyo 113-0033, Japan}
\author{Masahito Ueda}
\affiliation{Department of Physics, University of Tokyo, 7-3-1 Hongo, Bunkyo-ku, Tokyo 113-0033, Japan}
\affiliation{RIKEN Center for Emergent Matter Science (CEMS), Wako, Saitama 351-0198, Japan}
\date{\today}
%\keywords{}

% 03.75.Hh Static properties of condensates; thermodynamical, statistical, and structural properties
% 03.75.Lm Tunneling, Josephson effect, Bose-Einstein condensates in periodic potentials, solitons, vortices, and topological excitations 
% (see also 74.50.+r Tunneling phenomena; Josephson effects in superconductivity)
% 03.75.Mn Multicomponent condensates; spinor condensates
% 05.30.Jp Boson systems 
% (for static and dynamic properties of Bose-Einstein condensates, see 03.75.Hh and 03.75.Kk; see also 67.10.Ba Boson degeneracy in quantum fluids)
% 67.85.-d	Ultracold gases, trapped gases (see also 03.75.-b Matter waves in quantum mechanics)
% 67.85.Fg	Multicomponent condensates; spinor condensates
% 72.25.-b Spin polarized transport
% 73.43.-f Quantum Hall effects
% 73.43.Cd Theory and modeling
% 73.43.Nq Quantum phase transitions
%85.75.-d Magnetoelectronics; spintronics: devices exploiting spin polarized transport or integrated magnetic fields

%%%%%%%%%%%%%%%%%%%%%%%%%%%%%%%%%%%%%%%%%%%%%%%%%
% Abstract
%%%%%%%%%%%%%%%%%%%%%%%%%%%%%%%%%%%%%%%%%%%%%%%%%
\begin{abstract}
We study the ground-state phase diagram of two-dimensional two-component (or pseudospin-$\frac12$) Bose gases in a high synthetic magnetic field
in the space of the total filling factor and the ratio of the intercomponent coupling $g_{\ua\da}$ to the intracomponent one $g>0$. 
Using exact diagonalization, we find that when the intercomponent coupling is attractive ($g_{\ua\da}<0$), 
the product states of a pair of nearly uncorrelated quantum Hall states are remarkably robust and persist even when $|g_{\ua\da}|$ is close to $g$.  
This contrasts with the case of an intercomponent repulsion, 
where a variety of spin-singlet quantum Hall states with high intercomponent entanglement emerge for $g_{\ua\da}\approx g$. 
We interpret this marked dependence on the sign of $g_{\ua\da}$ in light of pseudopotentials on a sphere, 
and also explain recent numerical results in two-component Bose gases in mutually antiparallel magnetic fields 
where a qualitatively opposite dependence on the sign of $g_{\ua\da}$ is found. 
Our results thus unveil an intriguing connection between multicomponent quantum Hall systems and quantum spin Hall systems in minimal setups. 
\end{abstract}
\maketitle

%%%%%%%%%%%%%%%%%%%%%%%%%%%%%%%%%%%%%%%%%%%%%%%%%
% Main text
%%%%%%%%%%%%%%%%%%%%%%%%%%%%%%%%%%%%%%%%%%%%%%%%%

%%%%%%%%%%%%%%%%%%%%%%%%%%%%%%%%%%%%%%%%%%%%%%%%%
\section{Introduction}
%%%%%%%%%%%%%%%%%%%%%%%%%%%%%%%%%%%%%%%%%%%%%%%%%

% [ Synthetic gauge fileds ]--------------------
Engineering synthetic gauge fields in ultracold atomic systems has been a subject of active interest recently \cite{Dalibard11,Goldman14,Aidelsburger18}. 
While a real magnetic field does not produce a Lorentz force for neutral atoms, 
different methods of creating synthetic magnetic fields that do produce such a force have been developed. 
Such methods include mechanical rotation \cite{Madison00,AboShaeer01,Schweikhard04_LLL,Cooper08_review,Fetter09} and optical dressing \cite{Lin09} of atoms in continuum 
and laser-induced tunneling in optical lattices in real \cite{Aidelsburger13,Miyake13,Aidelsburger15} 
and synthetic \cite{Celi14,Mancini15,Stuhl15} spaces. 
For two-component (or pseudospin-$\frac12$) gases, which are populated in two hyperfine spin states of the same atomic species, 
a richer variety of gauge fields have been created, such as a uniform magnetic field by rotation \cite{Hall98,Schweikhard04_2comp}, 
and spin-orbit couplings \cite{Lin11,Wang12_SOC,Huang16,WuPan16,Zhai12,Galitski13} 
and pseudospin-dependent antiparallel magnetic fields \cite{Beeler13} by optical dressing techniques. 
By using these techniques, we can expect to emulate quantum Hall (QH) states 
and other topological states of matter in highly controlled atomic systems
and to explore many-body phenomena beyond the scope of other condensed matter systems \cite{Goldman16,Bloch12}. 
The capability to prepare {\it bosonic} particles in gauge fields is particularly unique to atomic systems. 
For moderate synthetic magnetic fields, a scalar Bose-Einstein condensate exhibits Abrikosov's triangular vortex lattice 
as observed experimentally \cite{AboShaeer01,Schweikhard04_LLL}. 
For high synthetic fields, theory predicts that the vortex lattice melts and 
that incompressible QH states appear at various integer and fractional values of the filling factor $\nu=N/N_\phi$,  
where $N$ is the number of atoms and $N_\phi$ is the number of flux quanta piercing the system \cite{Cooper08_review}. 
Such QH states of two-dimensional (2D) scalar Bose gases include a bosonic Laughlin state at $\nu=1/2$ \cite{Laughlin83,Wilkin98}, 
Jain's composite fermion (CF) states at $\nu=p/(p+1)~(p=2,3,\dots)$ \cite{Jain89,Regnault04,Chang05}, and a non-Abelian Moore-Read state at $\nu=1$ \cite{Moore91,Cooper01}. 
The Laughlin and Moore-Read states are two members of the Read-Rezayi series of states with an SU$(2)_k$ symmetry at $\nu=k/2~(k=1,2,\dots)$ \cite{Read99}. 

%############################
\begin{figure}
\begin{center}
\includegraphics[width=0.5\textwidth]{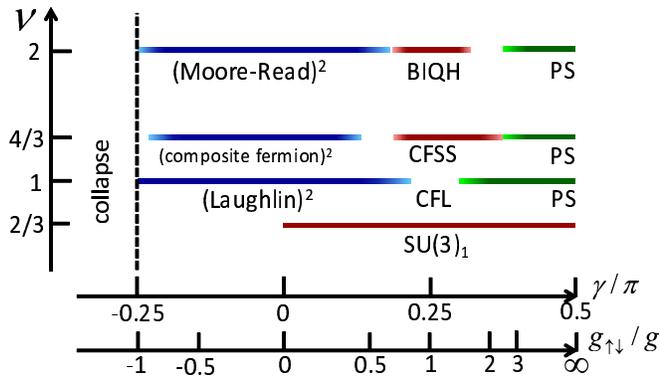}%{phase_qh2.eps}
\end{center}
\caption{(color online) 
GS phase diagram in the space of the total filling factor $\nu$ 
and the ratio $g_{\ua\da}/g=\tan\gamma$ of the intercomponent to intracomponent coupling constants. 
We assume an intracomponent repulsion $g>0$. 
For an intercomponent attraction $g_{\ua\da}<0$, 
the product states of a pair of nearly uncorrelated QH states (Laughlin, CF, and Moore-Read states) are found to appear over wide ranges of $g_{\ua\da}/g$. 
For $g_{\ua\da}\approx g$, in contrast, a variety of spin-singlet QH states with high intercomponent entanglement appear, 
such as the Halperin $(221)$ state at $\nu=2/3$ \cite{Halperin83,Paredes02}, a CFSS state at $\nu=4/3$ \cite{Wu13,Geraedts17}, 
%such as the SU$(2)_k$ states at $\nu=2k/3~(k=1,2)$ \cite{Halperin83,Paredes02,Ardonne99,Hormozi12,Grass12,FurukawaUeda12} 
and a BIQH state at $\nu=2$ \cite{Senthil13,FurukawaUeda13,Wu13,Regnault13,Grass14,Nakagawa17}. 
%(for $\nu=4/3$, a different scenario in favor of a composite fermion spin-singlet state has been given in Ref.\ \cite{Wu13}). 
Furthermore, a gapless spin-singlet CFL appears at $\nu=1$ \cite{Wu15}. 
For larger $g_{\ua\da}/g$, a phase separation (PS) occurs. 
The ranges of different phases indicated by shaded bars are determined in Sec.\ \ref{sec:ED}, 
and may contain errors due to finite-size effects or ambiguity in setting the condition for detecting the phase boundaries. 
}
\label{fig:phase}
\end{figure}
%############################

% [ Two-component systems ]--------------------
A large number of theoretical studies have recently been conducted 
for 2D pseudospin-$\frac12$ Bose gases in a uniform synthetic magnetic field, 
where richer physics than the scalar case is naturally expected. 
We introduce the total filling factor $\nu=(N_\ua+N_\da)/N_\phi$, 
where $N_\ua$ and $N_\da$ are the numbers of pseudospin-$\ua$ and $\da$ bosons, respectively. 
Within the Gross-Pitaevskii mean field theory which is valid for $\nu\gg 1$, several different types of vortex lattices have been shown to appear 
as the ratio of the intercomponent contact interaction $g_{\ua\da}$ to the intracomponent one $g>0$ is varied \cite{Mueller02,Kasamatsu03}. 
Meanwhile, studies on a high-magnetic-field regime with $\nu=O(1)$ have revealed 
that various spin-singlet QH states with a finite excitation gap emerge for pseudospin-independent (SU(2)-symmetric) interactions with $g_{\ua\da}=g>0$. 
Among those states, relatively large gaps are found for 
the Halperin $(221)$ state with an SU$(3)_1$ symmetry at $\nu=2/3$ \cite{Halperin83,Paredes02} 
%a non-Abelian SU$(3)_2$ state at $\nu=4/3$ \cite{Ardonne99,Hormozi12,Grass12,FurukawaUeda12}, 
and a bosonic integer QH (BIQH) state protected by a $U(1)$ symmetry at $\nu=2$ \cite{Senthil13,FurukawaUeda13,Wu13,Regnault13,Grass14,Nakagawa17,Geraedts17} 
(similar states have also been shown to appear in interacting scalar bosons in topological flat bands with Chern number two 
\cite{Barkeshli12,Wang12,Yang12,Liu12,Sterdyniak13,WuYL13,Sterdyniak15,Zeng16}, a correlated honeycomb lattice model \cite{He15}, 
and two-component bosons in topological flat bands \cite{Zeng17}). 
At $\nu=4/3$, two types of spin-singlet QH states compete in finite-size systems: 
a non-Abelian SU$(3)_2$ state \cite{Ardonne99,Hormozi12,Grass12,FurukawaUeda12} 
and a CF spin-singlet (CFSS) state \cite{Wu13}, 
with the latter selected in the thermodynamic limit \cite{Geraedts17}. 
Furthermore, a gapless spin-singlet composite Fermi liquid (CFL) has been shown to appear at $\nu=1$ \cite{Wu15,Geraedts17} 
(with an emergent particle-hole symmetry around this filling factor \cite{Geraedts17,Wang16,Mross16}). 
In all these spin-singlet states, the two components are highly entangled. 
For small $|g_{\ua\da}|/g$, in contrast, the system can be viewed as two weakly coupled scalar Bose gases, 
and the product states of nearly independent QH states (referred to as {\it doubled} QH states hereafter) are expected to appear. 
It is thus interesting to investigate the phase diagram with varying $g_{\ua\da}/g$ 
and analyze the competition among various QH states. 

% [ This paper ]--------------------
In this paper, we determine the ground-state (GS) phase diagram of pseudospin-$\frac12$ Bose gases in a uniform synthetic magnetic field 
in the space of the total filling factor $\nu$ and the coupling ratio $g_{\ua\da}/g$. 
To this end, we have performed an extensive exact diagonalization analysis in the lowest-Landau-level (LLL) basis on spherical and torus geometries. 
Our main results are summarized in Fig.\ \ref{fig:phase}. 
Here we parametrize the two coupling constants as
\begin{equation}\label{eq:gg_G}
 (g,g_{\ua\da})=G\ell^2 (\cos\gamma,\sin\gamma)
\end{equation}
with $G>0$, and change $\gamma$ in the range $-\pi/2 \le \gamma\le \pi/2$. 
As seen in this diagram, when the intercomponent coupling is attractive ($g_{\ua\da}<0$), 
doubled QH states are remarkably robust 
and persist even when $|g_{\ua\da}|$ is comparable to the intracomponent coupling $g>0$. 
This sharply contrasts with the case of an intercomponent repulsion $(g_{\ua\da}>0)$, 
where a variety of spin-singlet QH states with high intercomponent entanglement emerge for $g_{\ua\da}\approx g$. 
We interpret this remarkable dependence on the sign of $g_{\ua\da}$ in light of Haldane's pseudopotentials on a sphere \cite{Haldane83,Fano86}. 
More specifically, the stability of the doubled QH states for $g_{\ua\da}<0$ 
can be understood from the ``ferromagnetic'' nature of the intercomponent interaction in terms of (modified) angular momenta of particles. 
We note that some previous numerical works have also investigated the phase diagram in the space of the coupling ratio $g_{\ua\da}/g$ 
for $\nu=1$ \cite{Liu16}, $\nu=4/3$ \cite{FurukawaUeda12}, and $\nu=2$ \cite{Regnault13}. 
However, since these works set different conditions in determining the phase boundaries and might involve finite-size effects in different manners, 
it is worthwhile to reexamine the phase diagrams at these filling factors in a comparative manner along the same line of analyses. 
Furthermore, the case of $g_{\ua\da}<0$ was not analyzed in these works. 

% [ Comparison with the case of antiparallel fields ]--------------------
It is interesting to compare Fig.\ \ref{fig:phase} with 
the phase diagram of two-component Bose gases in {\it antiparallel} magnetic fields studied previously \cite{FurukawaUeda14} 
(see also Refs.\ \cite{Liu09,Fialko14} for earlier studies on the same and related systems). 
In the latter case, the pseudospin-$\ua$ ($\da$) component is subject to the magnetic field $+B$ ($-B$) in the direction perpendicular to the 2D gas, 
and the system possesses the time-reversal symmetry. 
Within the Gross-Pitaevskii mean-field theory which is valid for $\nu\gg 1$, 
one can show that the system in antiparallel fields shows the same vortex structures as the system in parallel fields studied in Refs.\ \cite{Mueller02,Kasamatsu03}. 
However, a remarkable distinction emerges in a high-field regime with $\nu=O(1)$: in the case of antiparallel fields, 
(fractional) quantum spin Hall states \cite{Bernevig06} composed of a pair of QH states with opposite chiralities 
are robust for an intercomponent repulsion $g_{\ua\da}>0$ and persist for $g_{\ua\da}$ as large as $g$. 
Similar results have also been found in the stability of two coupled bosonic Laughlin states in lattice models \cite{Repellin14}. 
These results suggest that the case of $g_{\ua\da}>0$ for antiparallel fields 
essentially corresponds to the case of $g_{\ua\da}<0$ for parallel fields. 
As discussed later, the pseudopotential approach also provides an insight into this intriguing correspondence. 

% [ Organization of the paper ]--------------------
The rest of the paper is organized as follows. 
In Sec.~\ref{sec:ED}, we present our exact diagonalization results. 
In particular, we perform an extensive search for incompressible states in the present system, 
and determine the ranges of different QH states shown in Fig.~\ref{fig:phase}. 
In Sec.~\ref{sec:interpret}, we discuss the stability of coupled QH states in light of pseudopotentials on a sphere. 
In Sec.~\ref{sec:summary}, we present a summary and an outlook for future studies. 
In Appendix \ref{app:QH_wvfn}, we summarize QH wave functions discussed in the paper. 
In Appendix \ref{app:pseudopot}, we describe some details on the calculation of pseudopotentials for two-component gases in antiparallel fields. 

%%%%%%%%%%%%%%%%%%%%%%%%%%%%%%%%%%%%%%%%%%%%%%%%%
\section{Exact diagonalization analysis}\label{sec:ED}
%%%%%%%%%%%%%%%%%%%%%%%%%%%%%%%%%%%%%%%%%%%%%%%%%

% [ System ]--------------------
In this section, we present our exact diagonalization analysis that has led to the phase diagram in Fig.~\ref{fig:phase}. 
We consider a system of a 2D pseudospin-$\frac12$ Bose gas (in the $xy$ plane) 
having two hyperfine spin states (labeled by $\alpha=\ua,\da$)
and subject to a synthetic magnetic field $B$ along the $z$ axis. 
In the case of a rotating gas, the field $B=2M\Omega/q$ is induced in the rotating frame of reference, 
where $M$ and $q$ are the mass and the fictitious charge, respectively, of a neutral atom and $\Omega$ is the rotation frequency. 
We denote the strengths of the intracomponent and intercomponent contact interactions by $g$ and $g_{\ua\da}$, respectively. 
In the second-quantized form, the interaction Hamiltonian is written as
\begin{equation}\label{eq:Hint}
 H_\mathrm{int} 
 = \sum_{\alpha,\beta=\ua,\da} \frac{g_{\alpha\beta}}{2} \int d^2\bm{r} \Psih_\alpha^\dagger(\rv) \Psih_\beta^\dagger(\rv) \Psih_\beta(\rv) \Psih_\alpha(\rv),
\end{equation}
where $\Psih_\alpha (\rv)$ is the bosonic field operator for the spin state $\alpha$. 
We set $g_{\ua\ua}=g_{\da\da}\equiv g>0$ and $g_{\ua\da}=g_{\da\ua}$.
For a 2D system of area $A$, the number of magnetic flux quanta piercing the system is given by 
$N_\phi=A/(2\pi \ell^2)$, where $\ell=\sqrt{\hbar/|qB|}$ is the magnetic length. 
Strongly correlated physics is expected to emerge when $N_\phi$ becomes comparable with or larger than 
the total number of particles, $N=N_\ua+N_\da$, for sufficiently high $B$. 
For such high $B$, it is useful to restrict ourselves to the low-energy subspace spanned by the LLL states. 
Within this restricted subspace, we have performed an exact diagonalization analysis of the interaction Hamiltonian \eqref{eq:Hint}. 
Our analysis presented here is quite analogous to the one performed for the systems in antiparallel fields in Ref.~\cite{FurukawaUeda14}. 

% We assume that $B$ is so large that the interaction energy is much smaller than the Landau-level spacing $\hbar |qB|/M$. 
%We investigate the GS phase diagram of the system in the space of the total filling factor $\nu=N/N_\phi$ and the interaction ratio $g_{\ua\da}/g$. 

%************************************************
\subsection{Spherical and torus geometries} \label{sec:geometries}
%************************************************

% [ Closed geometries ]--------------------
To study bulk properties, it is useful to work on closed uniform manifolds having no edge. 
In our analysis, we employ spherical \cite{Haldane83,Fano86} and torus \cite{Yoshioka84,Haldane85} geometries 
as was done in previous studies on the same and related systems \cite{Hormozi12,Grass12,FurukawaUeda12,FurukawaUeda13,Wu13,Regnault13,Nakagawa17,Wu15,Liu16,Grass13,Wu16,Geraedts17}. 
These geometries can describe the central region of a trapped gas, where the particle density is approximately uniform. 
Here we briefly describe the basic features of these geometries. 

% [ Spherical geometry ]--------------------
For a spherical geometry, a magnetic monopole of charge $- N_\phi (2\pi \hbar/q)$ with integer $N_\phi\equiv 2S$ is placed at the origin. 
It produces a uniform magnetic field $-B$ on the sphere of radius $R=\ell \sqrt{S}$. 
The LLL on a sphere corresponds to the subspace in which a certain modified angular momentum 
[as shown in Eq.~\eqref{eq:ang_mom} in Appendix \ref{app:pseudopot}] has the magnitude $S$, 
and is thus $(2S+1)$-fold degenerate. 
Introducing the spherical coordinates $(\theta,\phi)$ and the spinor coordinates 
\begin{equation}\label{eq:uv}
 u= \cos (\theta/2) e^{i\phi/2},~
 v= \sin (\theta/2) e^{-i\phi/2}, 
\end{equation}
single-particle orbitals in the LLL are given by $\psi_m\propto u^{S+m} v^{S-m}$, where 
$m\in\{-S,-S+1,\dots,S\}$ is the $z$-component of the angular momentum. 
On a sphere, the interaction Hamiltonian \eqref{eq:Hint} in the LLL subspace can conveniently be represented 
in terms of pseudopotentials \cite{Haldane83,Fano86}, as explained in Sec.~\ref{sec:interpret}. 
Because of the spherical symmetry, many-body eigenstates can be classified by the total angular momentum $L$. 

% [ Torus geometry ]--------------------
A torus geometry is formed by a periodic rectangle of sides $L_x$ and $L_y$. 
The degeneracy in the LLL manifold is given by $N_\phi=L_xL_y/(2\pi \ell^2)$. 
In our analysis, we set $L_x=L_y$. 
The representation of the interaction Hamiltonian \eqref{eq:Hint} in the LLL basis on this geometry can be found in, e.g., Ref.~\cite{Nakagawa17}. 
The many-body eigenstates can be classified by the total pseudomomentum $\bm{K}=(K_x,K_y)=2\pi\hbar(m_x/L_x,m_y/L_y)$. 
When $(N_\phi,N)=(q,p)\bar{N}$ with $\bar{N}$ being the largest common divisor of $N_\phi$ and $N$, 
the two integers $m_x$ and $m_y$ can take $m_x\in\{0,1,\dots,q\bar{N}-1\}$ and $m_y\in\{0,1,\dots,\bar{N}-1\}$. 
Since eigenstates with $m_x$ and $m_x+\bar{N}$ are related by a translation in the $y$ direction, 
all the eigenenergies are $q$-fold degenerate \cite{Haldane85}. 

% [ Shift ]--------------------
On both the sphere and the torus, the filling factor in the thermodynamic limit is given by $\nu=N/N_\phi$. 
For incompressible states on finite spheres, however, 
the relation between $N$ and $N_\phi$ involves a characteristic shift $\delta$ as follows:
\begin{equation}\label{eq:N_Nphi}
 N = \nu (N_\phi+\delta), 
\end{equation}
where $\delta$ depends on individual candidate wave functions. 
Therefore, on a sphere, competing incompressible states leading to the same $\nu$ in the thermodynamic limit 
can be studied separately with different $(N_\phi, N)$ if they have different shifts. 
On a torus, there is no shift, and all candidates for the same $\nu$ compete in the same finite-size calculation. 
% While a torus can give less biased results, 
% a sphere is more useful in discussing relative stabilities of given candidates. 

%************************************************
\subsection{Numerical search for incompressible states} \label{sec:incompress}
%************************************************

%############################
\begin{figure*}
\begin{center}
\includegraphics[width=0.47\textwidth]{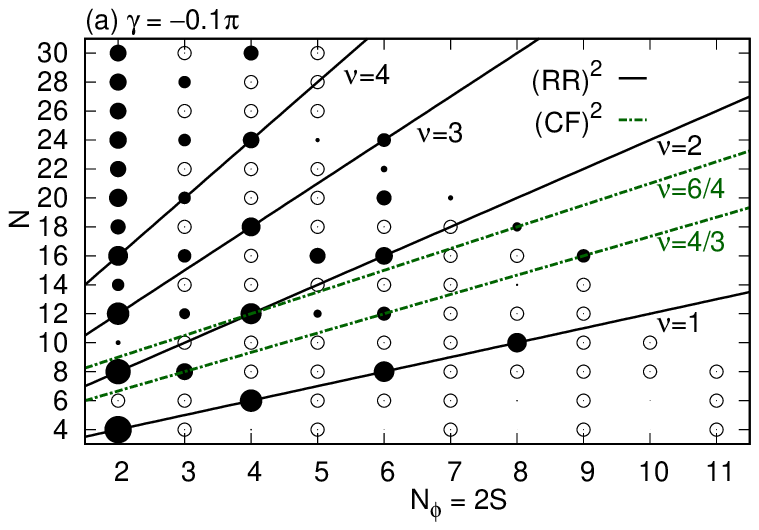} %{gsLqh_thm0p10.eps}
\includegraphics[width=0.47\textwidth]{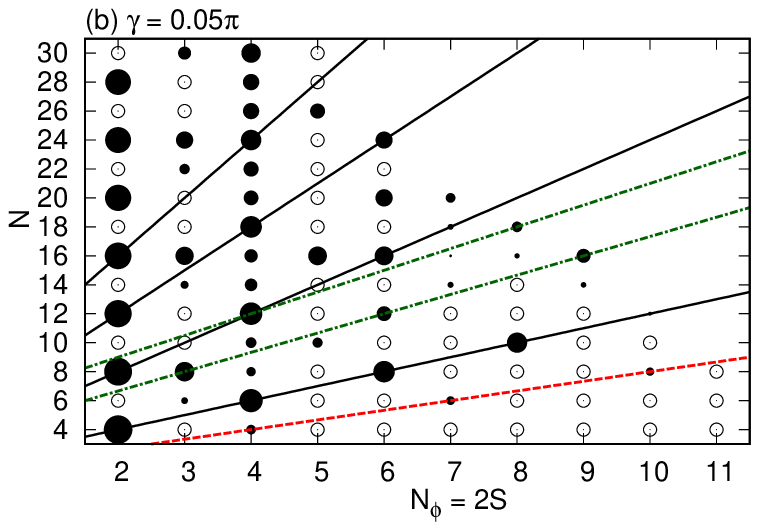} %{gsLqh_thp0p05.eps}
\includegraphics[width=0.47\textwidth]{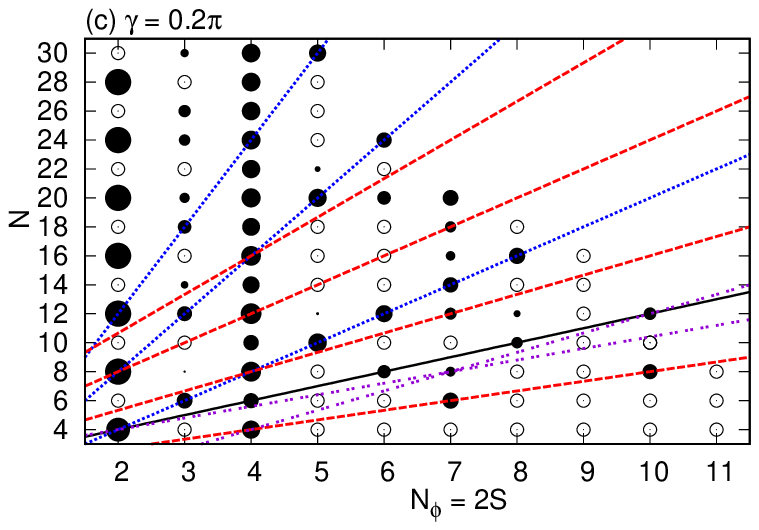} %{gsLqh_thp0p20.eps}
\includegraphics[width=0.47\textwidth]{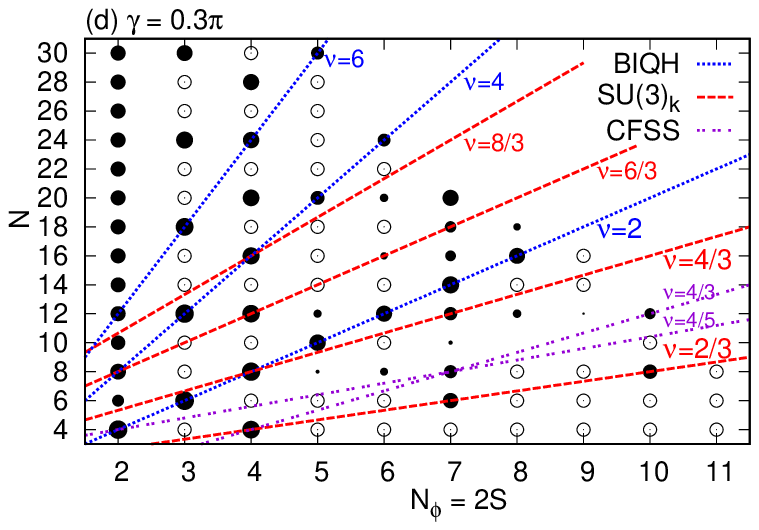} %{gsLqh_thp0p30.eps}
\end{center}
\caption{(color online) 
Candidates for incompressible GSs in the $(N_\phi,N)$ plane, 
calculated on a spherical geometry for different values of $\gamma=\arctan(g_{\ua\da}/g)$. 
Filled circles indicate GSs with the total angular momentum $L=0$, where incompressible states can appear; 
the area of each filled circle is proportional to the neutral gap $\Delta_n/G$. 
Empty circles indicate the GSs with $L>0$. 
Five types of lines indicate the relation \eqref{eq:N_Nphi} for different QH states:  
the doubled Read-Rezayi SU$(2)_k$ states [$(\nu,\delta)=(k,2)$; black solid], 
the doubled CF states [$(\nu,\delta)=(\frac{2p}{p+1},p+1)$; green dashed dotted], 
the SU$(3)_k$ states [$(\nu,\delta)=(\frac{2k}{3},2)$; red dashed], 
the BIQH state (with its possible generalizations) [$(\nu,\delta)=(2k,0)$; blue dotted], 
and the CFSS states [$(\nu,\delta)=(\frac45,3),(\frac43,-1)$; purple double dotted]. 
Data points are missing for large $N_\phi$ or $N$ due to an exponentially increasing computation time. 
}
\label{fig:gsL}
%Reference on the types of lines: http://dwg.jisw.com/01140/post_33.html
\end{figure*}
%############################

% [ Search for incompressible states ]--------------------
Through exact diagonalization calculations on a spherical geometry, 
we have carried out an extensive search for incompressible GSs in the $(N_\phi,N)$ plane 
for different values of $\gamma=\arctan(g_{\ua\da}/g)$ as shown in Fig.~\ref{fig:gsL}.  
Incompressible states, in general, appear as the unique GSs with $L=0$, which are indicated by filled circles. 
The area of each filled circle is proportional to the neutral gap $\Delta_n$ (in units of $G$ in Eq.~\eqref{eq:gg_G}), 
which is defined as the excitation gap for fixed $(N_\phi,N_\ua,N_\da)$. 
Five types of lines indicate the relation \eqref{eq:N_Nphi} for different candidate QH states; 
see Appendix \ref{app:QH_wvfn} for the wave functions of these states. 

% [ small \gamma ]--------------------
For small $\gamma$, doubled QH states are expected to appear. 
In Fig.~\ref{fig:gsL}, solid lines correspond to the doubled Read-Rezayi SU$(2)_k$ states at $\nu=k~(k=1,2,3,4)$, 
which include the doubled Laughlin ($k=1$) and Moore-Read $(k=2)$ states 
(we note that these states appear only for even $N_\phi$). 
Dashed dotted lines correspond to doubled CF states at $\nu=\frac{2p}{p+1} ~(p=2,3)$. 
For (a) $\gamma=-0.1\pi$ and (b) $\gamma=0.05\pi$, we find that $L=0$ GSs appear on these lines 
with relatively large excitation gaps $\Delta_n/G$ for $\nu=1$, $4/3$, and $2$. 
For $-\pi/2\lesssim \gamma\le 0$ (i.e., $-1\lesssim g_{\ua\da}/g\le 0$), we find that $L=0$ GSs continue to appear on these lines 
although the gaps $\Delta_n/G$ gradually shrink with increasing $|\gamma|$. 
In contrast, as we increase $\gamma$ in $0\le \gamma<\pi/4$, some of the GSs on these lines are replaced by $L>0$ states, as seen for (c) $\gamma=0.2\pi$. 
These results suggest that the doubled QH states are more stable for $\gamma<0$. 

% [ Around SU(2) point ]--------------------
At $\gamma=\pi/4$ (i.e., $g_{\ua\da}=g$), the system possesses the SU(2) spin rotational symmetry, 
and a variety of spin-singlet QH states appear as revealed in previous studies. 
Such spin-singlet QH states include the SU$(3)_k$ states at $\nu=2k/3~(k=1,2,\dots)$ 
\cite{Halperin83,Paredes02,Ardonne99,Hormozi12,Grass12,FurukawaUeda12}, 
a BIQH state at $\nu=2$ \cite{Senthil13,FurukawaUeda13,Wu13,Regnault13} (with possible generalizations to $\nu=4,6,\dots$ \cite{FurukawaUeda13}), 
and  CFSS states at $\nu=4/5$ and $4/3$ \cite{Wu13}. 
Since these states have finite excitation gaps, they are expected to be stable over some ranges around the SU(2) case. 
For (c) $\gamma=0.2\pi$ and (d) $\gamma=0.3\pi$ in Fig.~\ref{fig:gsL}, 
$L=0$ GSs are indeed found on the lines corresponding to these states
(for a similar plot in the SU(2) case $\gamma=\pi/4$, see Ref.~\cite{FurukawaUeda13}). 
In particular, relatively large gaps are found for the SU$(3)_1$ state at $\nu=2/3$ and the BIQH state at $\nu=2$. 
At $\nu=2/3$, the SU$(3)_1$ state (Halperin $(221)$ state) is known to be the exact zero-energy GS  for repulsive contact interactions $g,g_{\ua\da}>0$ \cite{Paredes02}. 
Although the BIQH and SU$(3)_3$ states compete at $\nu=2$, the gap for the former is (by a factor of about 1.5) larger than that for the latter, 
indicating that the BIQH state is likely to survive the competition \cite{FurukawaUeda13}.
At $\nu=4/3$, the SU$(3)_2$ and CFSS states compete; 
although the gap values for these states are close for the system sizes investigated in Fig.~\ref{fig:gsL}, 
a recent large-scale simulation based on the infinite density matrix renormalization group (iDMRG) has provided pieces of evidence that 
the CFSS state is stabilized in the thermodynamic limit \cite{Geraedts17}. 

%(see Ref.\ \cite{Wu13} for a different argument supporting the CFSS state). 

%############################
\begin{figure}
\begin{center}
\includegraphics[width=0.48\textwidth]{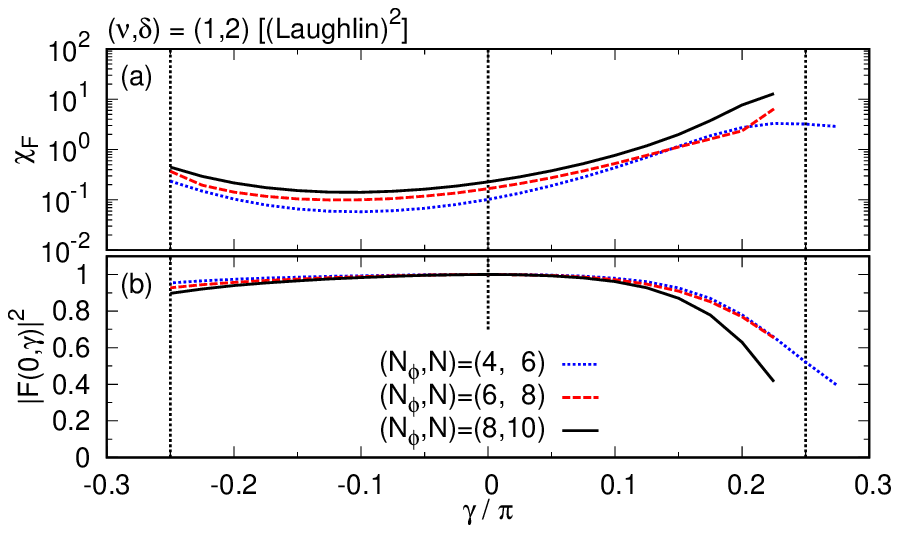} %{fidqh_ovlp_nut1.eps}
\includegraphics[width=0.48\textwidth]{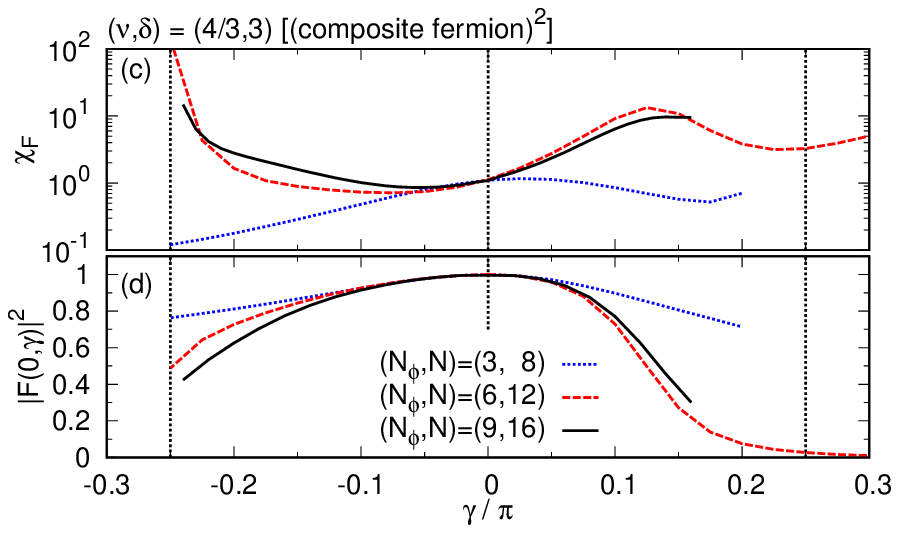} %{fidqh_ovlp_nut4o3.eps}
\includegraphics[width=0.48\textwidth]{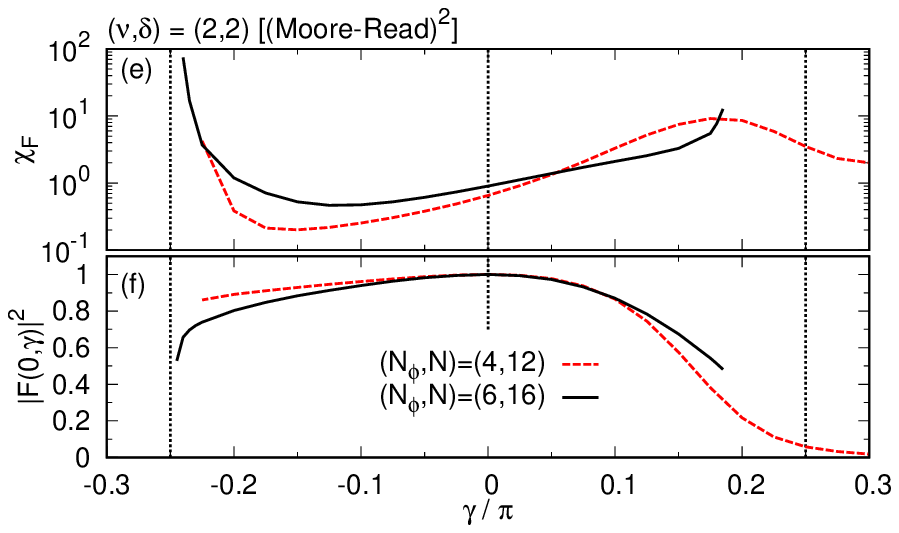} %{fidqh_ovlp_nut2.eps}
\end{center}
\caption{(color online) 
(a,c,e) The fidelity susceptibility $\chi_F(\gamma)$ and 
(b,d,f) the squared overlap $|F(0,\gamma)|^2$ with the decoupled case 
as functions of $\gamma$. 
Calculations are performed on a spherical geometry for $\nu=1$, $4/3$, and $2$. 
Curves are shown in the regions where the GS remains in the same sector. 
Vertical dotted lines correspond to $\gamma=0,\pm \pi/4$. 
}
\label{fig:fidqh_ovlp_g0}
\end{figure}
%############################

%############################
\begin{figure}
\begin{center}
\includegraphics[width=0.48\textwidth]{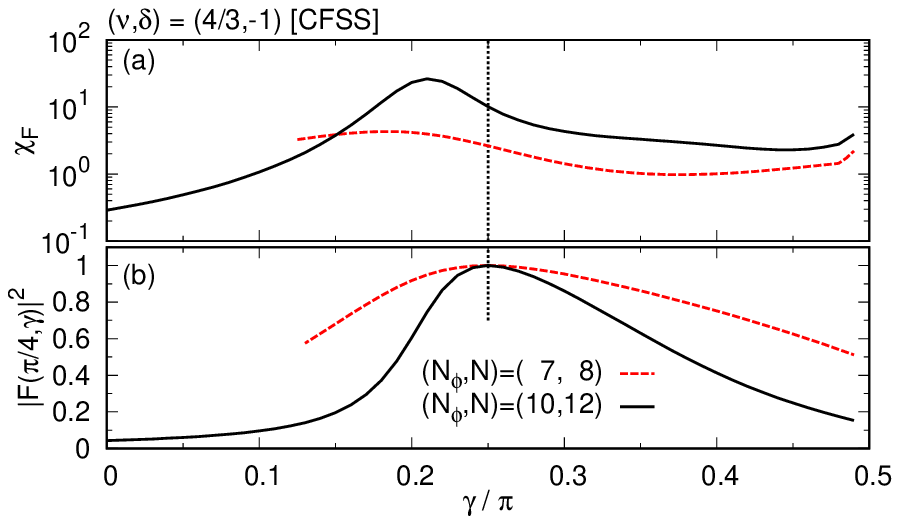} %{fidqh_ovlp_nut4o3d2.eps}
\includegraphics[width=0.48\textwidth]{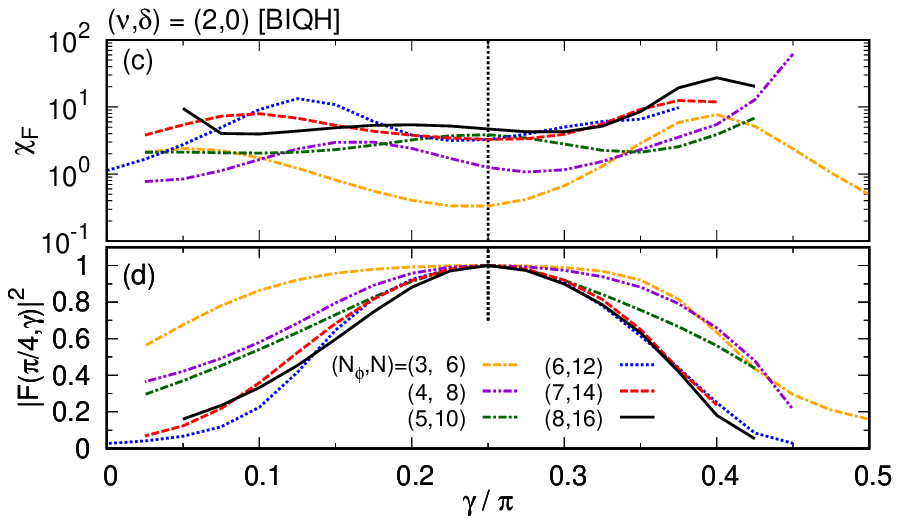} %{fidqh_ovlp_nut2d0.eps}
\end{center}
\caption{(color online) 
(a,c) The fidelity susceptibility $\chi_F(\gamma)$, 
(b) the squared overlap with the SU$(3)_2$ wave function, and 
(d) the squared overlap $|F(\pi/4,\gamma)|^2$ with the SU$(2)$ case, 
as functions of $\gamma$ around $\gamma=\pi/4$ (the vertical dotted lines). 
Calculations are performed on a spherical geometry. 
Curves are shown in the regions where the GS remains in the same sector. 
}
\label{fig:fidqh_ovlp_g1}
\end{figure}
%############################

% [ Continuation to the next subsection ]--------------------
This section has focused on a global picture of the types and the ranges of incompressible QH states present in the system. 
More precise estimation of the range of each QH phase requires a more detailed analysis, which we present in the next section. 
Before closing the section, we note that the appearance of the $L=0$ GS as examined here is not a sufficient condition 
for incompressibility---incompressibility is guaranteed by further showing the robustness of the excitation gap in the thermodynamic limit. 
However, since only a few system sizes are available for each candidate QH state in exact diagonalization,  
one cannot make a reliable extrapolation of the excitation gap to the thermodynamic limit. 
In the next section, we use different quantities (mainly, the overlap of the GS with a representative wave function) 
and the knowledge gained from a related system in antiparallel fields \cite{FurukawaUeda14} 
to estimate the range of each QH phase. 

%************************************************
\subsection{Ranges of quantum Hall states} \label{sec:determine_phase}
%************************************************

%############################
\begin{figure*}
\begin{center}
\includegraphics[width=0.48\textwidth]{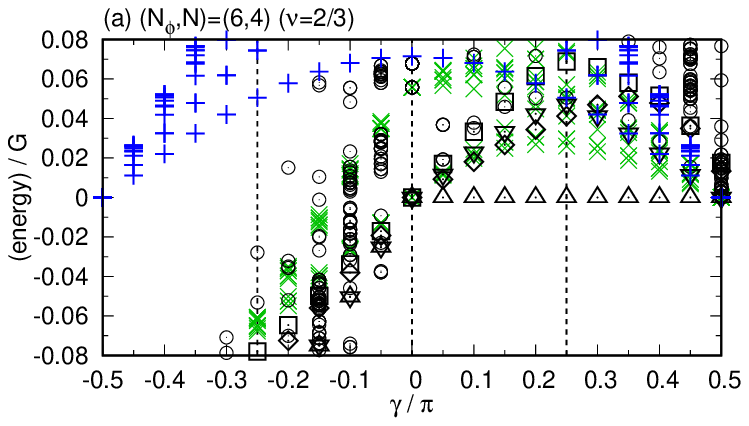} %{ener_gamma_Ns06N04.eps}
\includegraphics[width=0.48\textwidth]{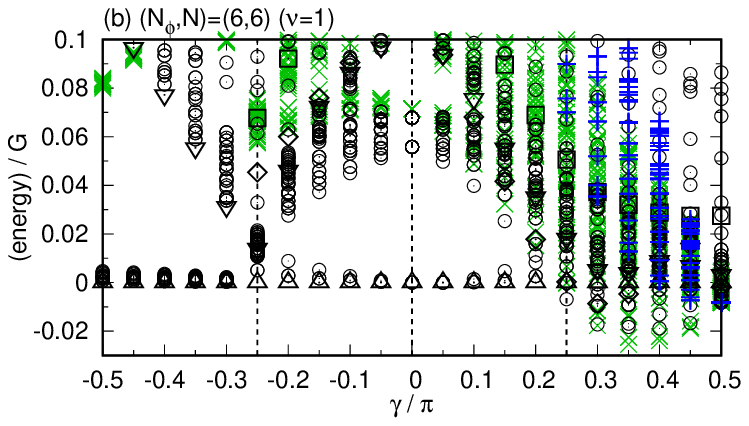} %{ener_gamma_Ns06N06.eps}
\includegraphics[width=0.48\textwidth]{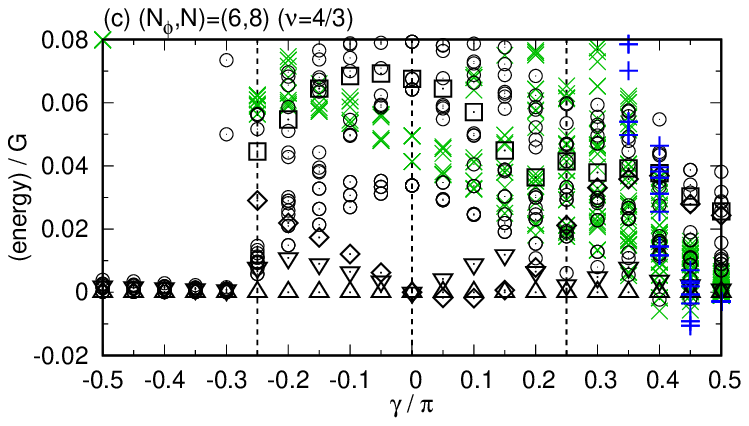} %{ener_gamma_Ns06N08.eps}
\includegraphics[width=0.48\textwidth]{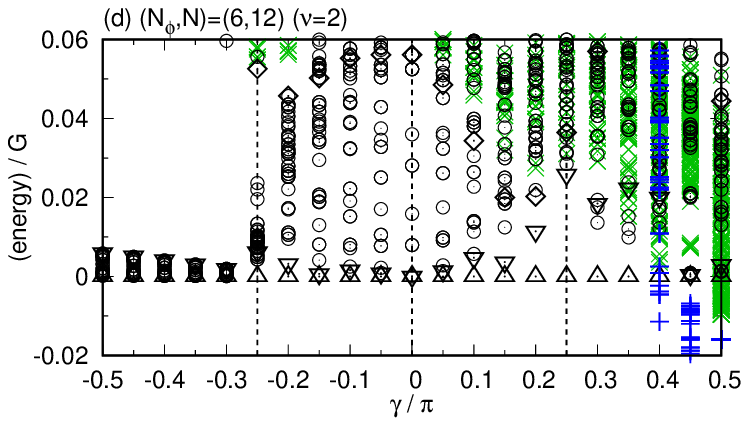} %{ener_gamma_Ns06N12.eps}
\end{center}
\caption{(color online) 
Energy spectra versus $\gamma$ for $\nu=2/3$, $1$, $4/3$, and $2$ with $N_\phi=6$ on a torus geometry. 
The eigenstates are classified by the magnetization (or population imbalance) $S_z=(N_\ua-N_\da)/2$ and the pseudomomentum $\Kv$. 
Upward ($\triangle$) and downward ($\triangledown$) triangles indicate the two lowest-energy states 
in the sector with $\Kv=\bm{0}$ and $\rho_\pi=p_{\ua\da}=+1$ in the equal-population case $S_z=0$, 
where $\rho_\pi$ and $p_{\ua\da}$ are the quantum numbers 
associated with the $\pi$ spatial rotation and the interchange of the two components, respectively. 
The lowest energy in this sector is subtracted from the entire spectrum in (b,c,d). 
Diamonds ($\diamond$) and squares ($\square$) indicate the two lowest-energy states in the sector with $\Kv=\bm{0}$ and $\rho_\pi=p_{\ua\da}=-1$. 
Circles ($\bigcirc$) indicate other eigenstates in the equal-population case. 
Greek ($+$) and diagonal ($\times$) crosses indicate eigenstates in the fully ($S_z=\pm N/2$) and partially ($0<|S_z|<N/2$) imbalanced cases, respectively. 
Only the two lowest energies are displayed in each sector. 
For $\nu=p/q$ (with $p$ and $q$ being coprime), each eigenenergy is $q$-fold degenerate. 
}
\label{fig:spec_torus}
\end{figure*}
%############################

% [ Subsection introduction ]--------------------
Hereafter we focus on the filling factors $\nu=2/3$, $1$, $4/3$, and $2$, where the QH states have relatively large excitation gaps. 
We determine the range of $\gamma$ over which each QH state identified in Sec.\ \ref{sec:incompress} is stabilized. 

% [ Fidelity susceptibility and squared overlap ]--------------------
Similar to Ref.~\cite{FurukawaUeda14}, we examine two kinds of quantities for this purpose: 
the fidelity susceptibility and the squared overlap of the GS with representative wave functions. %(see Figs.\ \ref{fig:fidqh_ovlp_g0} and \ref{fig:fidqh_ovlp_g1}). 
The fidelity susceptibility $\chi_F$ measures how fast the GS changes as a function of $\gamma$, and is defined as \cite{You07} 
\begin{equation} 
\chi_F(\gamma) = -2 \lim_{\delta\gamma\to 0} \frac{\ln F(\gamma,\gamma+\delta\gamma)}{(\delta\gamma)^2}, 
\end{equation}
where $F(\gamma,\gamma+\delta\gamma)=|\langle\Psi(\gamma)|\Psi(\gamma+\delta\gamma)\rangle|$ is 
the overlap between the GSs at two close points $\gamma$ and $\gamma+\delta\gamma$. 
A peak in this quantity, in general, signals a phase transition. 
This quantity has proven to be quite useful for detecting phase transitions in the case of antiparallel fields \cite{FurukawaUeda14}. 
In the present case of parallel fields, however, $\chi_F$ does not show a clear peak structure or a smooth dependence on the system size; 
this may be attributed to severer finite-size effects due to more complicated competition among various phases. 
Nonetheless, exact diagonalization is still useful in a regime where a certain QH state clearly wins for given $(N_\phi,N)$. 
The squared overlap of the GS with a representative wave function can be used to identify such a regime. 
%Thus we mainly use the squared overlap of the GS with representative wave functions in the estimation of phase boundaries. 

%We thus perform a more elaborate analysis in combination with the behavior of the squared overlap. {\bf (better expression?)}

% [ Doubled QH states ]--------------------
In Fig.\ \ref{fig:fidqh_ovlp_g0}, we analyze the ranges of the doubled QH states. 
In the decoupled case ($\gamma=0$), the GS for $(\nu,\delta)=(1,2)$ is given exactly by the doubled Laughlin wave functions \cite{Wilkin98}. 
At the same point, the GSs for $(\nu,\delta)=(4/3,3)$ and $(2,2)$ have large overlaps with the doubled CF wave functions 
and the doubled Moore-Read wave functions, respectively; 
indeed, the squared overlaps with these wave functions are $0.98^2$ and $0.96687^2$ for $(N_\phi,N)=(9,16)$ and $(6,16)$, respectively \cite{Chang05}, 
where the square is due to the presence of two components. 
In Fig.\ \ref{fig:fidqh_ovlp_g0}(b,d,f), we plot the squared overlap $|F(0,\gamma)|^2$ of the GS with the decoupled case ($\gamma=0$)
to analyze the stability of these doubled QH states. 
We find that $|F(0,\gamma)|^2$ decreases more slowly for $\gamma<0$ than for $\gamma>0$ as we move away from the decoupled case; 
this indicates that the doubled QH states are more robust for an intercomponent attraction $g_{\ua\da}<0$. 
In general, the squared overlap can only show a smooth behavior across a phase transition point in finite-size systems (unless the GS moves to another sector of the Hilbert space); 
furthermore, it tends to decrease exponentially with the system size owing to an exponentially increasing Hilbert space dimension. 
To estimate the ranges of the doubled QH states from the present data, 
a useful guidance can be gained from Ref.\ \cite{FurukawaUeda14}: in the case of antiparallel fields, 
a peak in the fidelity susceptibility $\chi_F$ is found when the squared overlap $|F(0,\gamma)|^2$ is around $0.5$ 
for the largest system size treated in each of Fig.~\ref{fig:fidqh_ovlp_g0}(b,d,f). 
Using the data for such system sizes and finding the points where $|F(0,\gamma)|^2$ becomes $0.5$ or the GS moves to another total-angular-momentum sector, 
we can estimate the ranges of the doubled QH states as follows: 
\begin{equation}\label{eq:range_dqh}
\begin{split}
 \text{(Laughlin)}^2: &-0.25\pi<\gamma\lesssim 0.22\pi; \\
 \text{(composite~fermion)}^2: &-0.23\pi\lesssim\gamma\lesssim 0.13\pi; \\
 \text{(Moore-Read)}^2: &-0.25\pi \lesssim\gamma\lesssim 0.18\pi. 
\end{split}
\end{equation}
Around the boundaries of these ranges, $\chi_F$ shows peaks or takes relatively large values as seen in Fig.~\ref{fig:fidqh_ovlp_g0}(a,c,e).  
We note that the above estimates can contain errors due to finite-size effects or ambiguity in setting the condition for $|F(0,\gamma)|^2$. 
A more precise determination of phase boundaries requires a simulation for larger systems 
by using, e.g., the DMRG \cite{Shibata01,Feiguin08,Kovrizhin10,Geraedts17}. 

% [ Spin-singlet QH states ]--------------------
% In Fig.\ \ref{fig:fidqh_ovlp_g1}(b), we examine the squared overlap of the GS with the SU$(3)_2$ wave function for $(\nu,\delta)=(4/3,2)$; 
% the latter is constructed by calculating the GS for the 3-body contact interaction [see Eq.~\eqref{eq:H_k} in Appendix \ref{app:QH_wvfn}]. 
% Finding the points where $|\bracket{\text{SU}(3)_2}{\Psi(\gamma)}|^2=0.5$ or the GS sector changes in the data for $(N_\phi,N)=(7,12)$, 
% we estimate the range of the SU$(3)_2$ state to be $0.14\pi\lesssim\gamma\lesssim 0.36\pi$. 

We have performed a similar analysis to estimate the ranges of the spin-singlet QH states as shown in Fig.\ \ref{fig:fidqh_ovlp_g1}. 
In Fig.\ \ref{fig:fidqh_ovlp_g1}(b), we examine the squared overlap $|F(\pi/4,\gamma)|^2$ with the SU$(2)$ case ($\gamma=\pi/4$) for $(\nu,\delta)=(4/3,-1)$
to analyze the range of the CFSS state. 
We note that the squared overlap between our reference state $\ket{\Psi(\pi/4)}$ and the CFSS wave function 
is $0.9711$, a value close to unity, for $(N_\phi,N)=(10,12)$ \cite{Geraedts17}. 
Finding the points where $|F(\pi/4,\gamma)|^2=0.5$ for $(N_\phi,N)=(10,12)$, 
we estimate the range of the CFSS state to be $0.19\pi\lesssim\gamma\lesssim 0.38\pi$ \cite{Comment_CFSS}. 
In Fig.\ \ref{fig:fidqh_ovlp_g1}(d), we examine $|F(\pi/4,\gamma)|^2$ for $(\nu,\delta)=(2,0)$ to analyze the range of the BIQH state. 
Here, the squared overlap between our reference state $\ket{\Psi(\pi/4)}$ and the BIQH wave function is $0.8197$ for $(N_\phi,N)=(8,16)$ \cite{Geraedts17}. 
The condition $|F(\pi/4,\gamma)|^2=0.5$ leads to the range $0.13\pi\lesssim\gamma\lesssim 0.37\pi$,  
which overlaps with the estimated range of the doubled Moore-Read states in Eq.\ \eqref{eq:range_dqh}. 
Since the overlap between $\ket{\Psi(\pi/4)}$ and the BIQH wave function is not very close to unity, we need a stricter condition. 
With the condition $|F(\pi/4,\gamma)|^2=0.8$, for example, 
we can estimate the range of the BIQH state to be $0.18\pi\lesssim\gamma\lesssim 0.32\pi$. 
An even stricter condition was used in Ref.~\cite{Regnault13}. 

%In Fig.\ \ref{fig:fidqh_ovlp_g1}(d), we examine the squared overlap $|F(\pi/4,\gamma)|^2$ with the SU$(2)$ case for $(\nu,\delta)=(2,0)$
%to analyze the range of the BIQH state. 

%############################
\begin{figure}
\begin{center}
\includegraphics[width=0.45\textwidth]{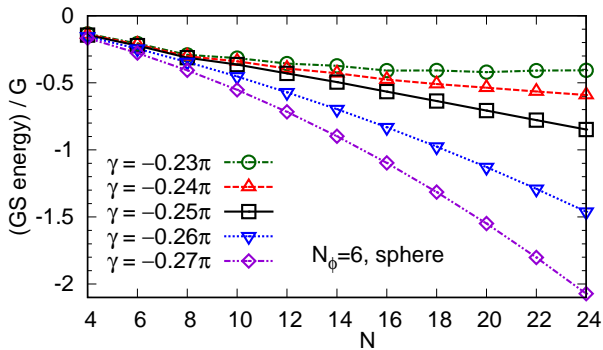}%{eneM6.eps}
\end{center}
\caption{(color online) 
GS energy as a function of $N$ for different values of $\gamma$ around $-\pi/4$. 
We use a spherical geometry and set $N_\phi=6$. 
}
\label{fig:energy_N}
\end{figure}
%############################

% [ Energy spectra ]--------------------
Finally, we examine energy spectra on a torus geometry in Fig.\ \ref{fig:spec_torus}. 
A torus geometry can provide less biased results since there is no shift and 
different candidates of QH states can compete in the same finite-size calculation. 
However, the presence of topological degeneracy can make the analysis more complex. 
For $\nu=2/3$ in Fig.\ \ref{fig:spec_torus}(a), 
we can clearly see the presence of a gap above the zero-energy SU$(3)_1$ state (with $3$-fold degeneracy) for $0<\gamma<\pi/2$. 
For $\nu=1$ in Fig.\ \ref{fig:spec_torus}(b), there appear $4$-fold degenerate zero-energy GSs at $\gamma=0$, 
which are given by the products of Laughlin states; 
a large gap opens above these GSs, and it decreases more slowly for $\gamma<0$ than for $\gamma>0$ with increasing $|\gamma|$.
Although the behaviors of the spectra are more complex for $\nu=4/3$ and $2$ as shown in Fig.\ \ref{fig:spec_torus}(c,d), 
we can see the emergence of energy gaps above the doubled CF states [around $\gamma=0$ in (c)]
and the BIQH state [around $\gamma=1$ in (d)]. 
In Fig.\ \ref{fig:spec_torus}(b,c,d), we can further find the occurrence of a phase separation for large $\gamma$ 
through the replacement of the GS with an imbalanced state with $N_\ua\ne N_\da$. 
The boundaries of phase separated regions in Fig.~\ref{fig:phase} are estimated in this way from Fig.\ \ref{fig:spec_torus}(b,c,d). 

%The gap above the SU$(3)_2$ state with $6$-fold degeneracy at $\nu=4/3$ is analyzed in more detail in Refs.~\cite{Grass12,FurukawaUeda12}. 

%************************************************
\subsection{Collapse of the gas for $\gamma<-\pi/4$} \label{sec:collapse}
%************************************************

Similar to the case of antiparallel fields \cite{FurukawaUeda14}, 
a collapse of the gas occurs for $\gamma<-\pi/4$ owing to the dominance of an intercomponent attraction. 
As seen in Fig.\ \ref{fig:energy_N}, the GS energy $E_\mathrm{GS}(N)$ as a function of $N$ 
is convex for $\gamma>-\pi/4$ and is concave for $\gamma<-\pi/4$. 
This indicates that the compressibility $\kappa$, which is inversely proportional to $\frac{d^2 E_\mathrm{GS}}{dN^2}$, 
changes its sign across $\gamma=-\pi/4$ (with a divergence $\kappa\to\pm\infty$ at the transition point). 
The state with $\kappa<0$ for $\gamma<-\pi/4$ is thermodynamically unstable and spontaneously contracts, 
leading to a collapse of the gas \cite{Comment_collapse}.

%%%%%%%%%%%%%%%%%%%%%%%%%%%%%%%%%%%%%%%%%%%%%%%%%
\section{Intercomponent entanglement and pseudopotentials}\label{sec:interpret}
%%%%%%%%%%%%%%%%%%%%%%%%%%%%%%%%%%%%%%%%%%%%%%%%%

% [ Section introduction: stability of doubled QH states ]--------------------
The phase diagram in Fig.\ \ref{fig:phase}, which is determined in the preceding section,  
shows a remarkable dependence on the sign of the intercomponent coupling $g_{\ua\da}$. 
While doubled QH states are robust for $g_{\ua\da}<0$, 
they are destabilized for moderate $g_{\ua\da}/g>0$, 
and a variety of spin-singlet QH states with high intercomponent entanglement emerge for $g_{\ua\da}/g\approx 1$. 
Interestingly, a qualitatively opposite dependence on the sign of $g_{\ua\da}$ has been found in two-component Bose gases in antiparallel fields \cite{FurukawaUeda14}; 
in this case, the products of a pair of QH states are more stable for $g_{\ua\da}>0$ than for $g_{\ua\da}<0$. 
In this section, we present an interpretation of these results  in light of pseudopotentials on a spherical geometry. 

% [ Pseudopotentials ]--------------------
The pseudopotential representation of interactions is introduced in the following way \cite{Haldane83,Fano86}. 
In a scattering process of two particles on a sphere, their total angular momentum is conserved because of the spherical spatial symmetry. 
The two-body interaction Hamiltonian \eqref{eq:Hint} can therefore be decomposed as
\begin{equation}\label{eq:H_VAA}
 H_\mathrm{int} = \frac12 \sum_{\alpha,\beta=\ua,\da} \sum_{J=0}^{2S} V_J^{\alpha\beta} \sum_{M=-J}^J A_{JM}^{\alpha\beta \dagger} A_{JM}^{\alpha\beta}. 
\end{equation}
Here, we have introduced the pair creation operator
\begin{equation}
 A_{JM}^{\alpha\beta \dagger} = \sum_{m_1+m_2=M} b_{m_1\alpha}^\dagger b_{m_2\beta}^\dagger \bracket{S,m_1;S,m_2}{J,M}, 
\end{equation}
where $b_{m\alpha}^\dagger$ is the bosonic creation operator for the pseudospin state $\alpha$ and the $m$-th orbital in the LLL, 
and $\bracket{S,m_1;S,m_2}{J,M}$ is the Clebsch-Gordan coefficient. 
We note that $A_{JM}^{\alpha\alpha}=0$ when $S-J$ is odd, owing to the bosonic statistics. 
The coefficient $V_J^{\alpha\beta}$ describes the interaction energy of two particles 
with pseudospin states $\alpha$ and $\beta$ in the total angular momentum $J$, and is called the pseudopotential. 
The expansion \eqref{eq:H_VAA} is analogous to the decomposition of an interaction between spinor atoms 
in terms of the total spin magnitude \cite{Kawaguchi12,Lian14}.

%############################
\begin{figure}
\begin{center}
\includegraphics[width=0.45\textwidth]{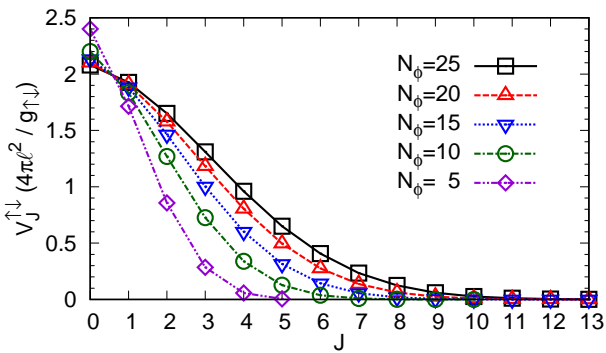}%{ppot_apara.eps}
\end{center}
\caption{(color online) 
Intercomponent pseudopotential \eqref{eq:Vud_J} as a function of 
the total angular momentum $J$ of two particles in antiparallel fields on a sphere. 
}
\label{fig:ppot_apara}
\end{figure}
%############################

% [ Expressions of pseudopotentials ]--------------------
In the case of two-component gases in parallel fields, the pseudopotentials are calculated to be
\begin{equation}\label{eq:Vab_J}
 V_J^{\alpha\beta}= \delta_{J,2S} \frac{g_{\alpha\beta}}{4\pi\ell^2} \frac{(2S+1)^2}{S(4S+1)}. 
\end{equation}
As seen in this expression, $V_J^{\alpha\beta}$ is nonzero only when $J$ takes the maximal value $2S$. 
In the case of two-component gases in antiparallel fields, the intracomponent pseudopotentials $V_J^{\alpha\alpha}$ are given by the same form as Eq.~\eqref{eq:Vab_J} 
while the intercomponent one is given by 
\begin{equation}\label{eq:Vud_J}
 V_J^{\ua\da} = \frac{g_{\ua\da}}{4\pi\ell^2} \frac{[(2S+1)!]^2}{S (2S-J)!(2S+J+1)!}.
\end{equation}
See Appendix \ref{app:pseudopot} for the derivation of Eqs.\ \eqref{eq:Vab_J} and \eqref{eq:Vud_J}. 
Equation \eqref{eq:Vud_J} is plotted in Fig.~\ref{fig:ppot_apara}. 
As seen in this figure, $V_J^{\ua\da}$ in units of $g_{\ua\da}/(4\pi\ell^2)$ takes the maximum of about $2$ for $J=0$, 
and decreases monotonically with increasing $J$; 
furthermore, with increasing $N_\phi=2S$, the decrease of $V_J^{\ua\da}$ as a function of $J$ becomes slower. 

%############################
\begin{figure}
\begin{center}
\includegraphics[width=0.5\textwidth]{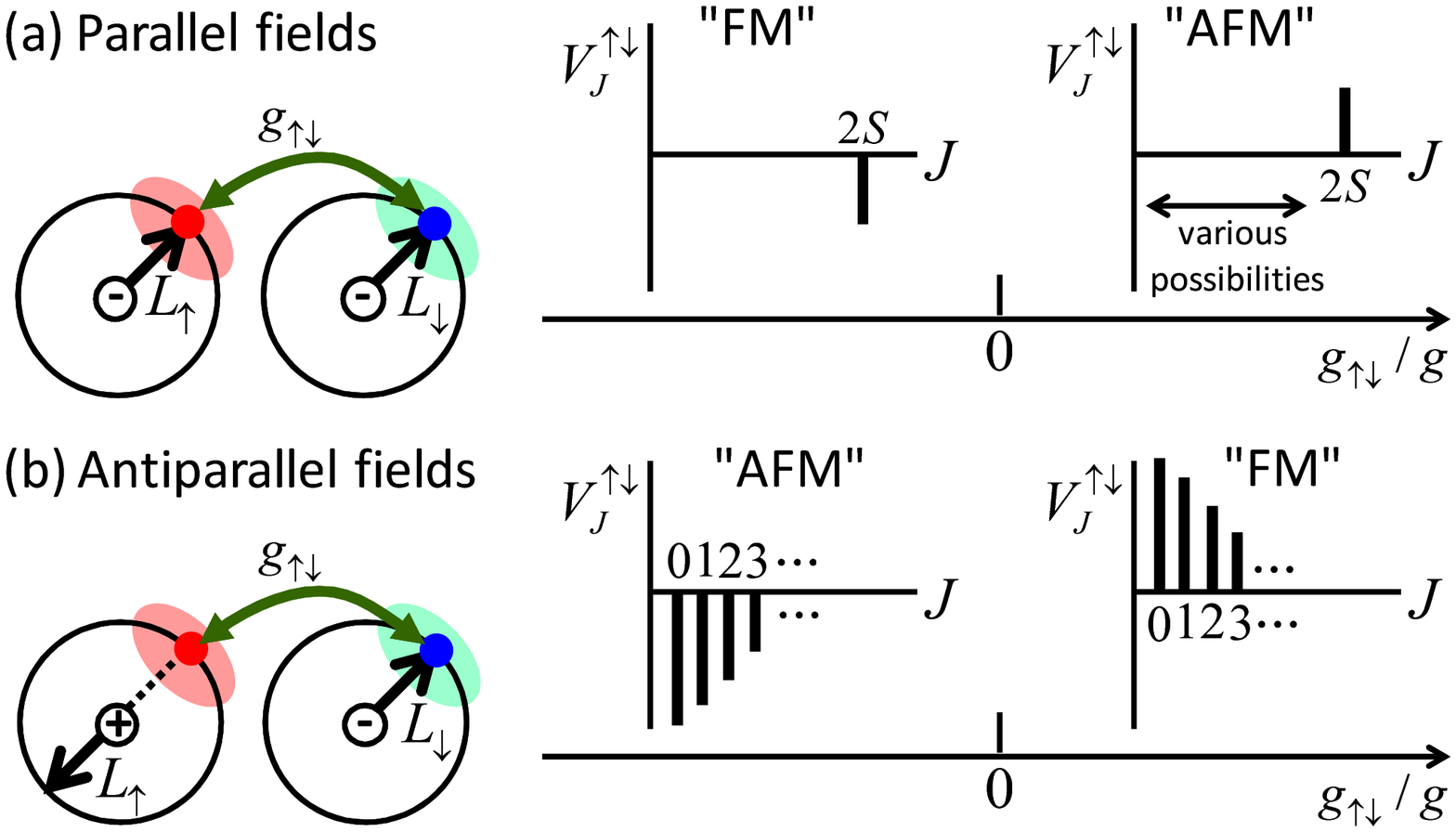}%{phase_ppot.eps}
\end{center}
\caption{(color online) 
Schematic pictures of modified angular momenta $\Lv_{\ua,\da}$ (left) and the intercomponent pseudopotential $V^{\ua\da}_J$ (right) on a spherical geometry. 
The sign at the center of a sphere indicates the sign of the magnetic monopole charge in units of $2\pi\hbar/q$ for each component.  
See Eqs.\ \eqref{eq:Vab_J} and \eqref{eq:Vud_J} for the expressions of $V_J^{\ua\da}$ 
and Fig.\ \ref{fig:ppot_apara} for the plot of Eq.\ \eqref{eq:Vud_J}. 
In the case of (a) parallel fields, 
a repulsive (attractive) coupling $g_{\ua\da}>0$ ($g_{\ua\da}<0$) between $\ua$ and $\da$ particles
can be viewed as an ``AFM'' (``FM'') interaction between their angular momenta. 
In the case of (b) antiparallel fields, in contrast, 
an intercomponent repulsion (attraction) can be viewed as a ``FM'' (``AFM'') interaction between the angular momenta. 
The intercomponent coupling of an ``AFM'' type is expected to produce higher entanglement between the two components than the one of a ``FM'' type. 
An intercomponent repulsion $g_{\ua\da}>0$ in (a) has a particularly large flexibility in the way of forming entanglement between the two components, 
which qualitatively explains why a variety of spin-singlet QH states emerge for $g_{\ua\da}\approx g$. 
}
\label{fig:phase_ppot}
\end{figure}
%############################

% [ Understanding of the behavior of pseudopotentials ]--------------------
Figure \ref{fig:phase_ppot} summarizes the behaviors of the intercomponent pseudopotential $V_J^{\ua\da}$ for parallel and antiparallel fields (right) 
and presents their interpretations in terms of angular momenta (left). 
In the case of (a) parallel fields, a particle is located around the direction of its angular momentum $\langle \Lv_\alpha \rangle$ \cite{Haldane83,Fano86}.  
In this case, a repulsive (attractive) interaction between $\ua$ and $\da$ particles
can be viewed as an ``antiferromagnetic (AFM)'' [``ferromagnetic (FM)''] interaction between their angular momenta $\Lv_{\ua,\da}$. 
This is consistent with the behavior of $V_J^{\ua\da}$ in Eq.\ \eqref{eq:Vab_J}, 
which disfavors (favors) the maximal total angular momentum $J=2S$ for $g_{\ua\da}>0$ ($g_{\ua\da}<0$). 
In the case of (b) antiparallel fields, in contrast, a pseudospin-$\ua$ particle is located 
around the direction of $-\langle \Lv_\ua \rangle$. 
Thus, a repulsive (attractive) interaction between $\ua$ and $\da$ particles can be viewed as 
a ``FM'' (``AFM'') interaction between their angular momenta $\Lv_{\ua,\da}$. 
This is consistent with Eq.\ \eqref{eq:Vud_J}, which disfavors (favors) states with small $J$ for $g_{\ua\da}>0$ ($g_{\ua\da}<0$). 

% [ Interpretation of the phase diagrams ]--------------------
Now the phase diagrams in the cases of parallel and antiparallel fields can be understood as follows. 
In the absence of an intercomponent coupling $g_{\ua\da}$, the intracomponent pseudopotential $V_J^{\alpha\alpha}$ having an ``AFM'' nature 
leads to the formation of QH states in each component. 
Such QH states reside in the singlet sector ($L=0$) of the total angular momentum, 
and thus are highly entangled with respect to angular momenta of particles in each component. 
In the case of parallel (antiparallel) fields, an intercomponent attraction $g_{\ua\da}<0$ (repulsion $g_{\ua\da}>0$) 
introduces a ``FM'' interaction between angular momenta of particles in different components. 
Since such an interaction favors the formation of product states such as $\ket{J=2S,M=2S}=\ket{S,S;S,S}$ 
for two particles in different components, 
it is not likely to produce high entanglement between the components. 
In contrast, an intercomponent repulsion $g_{\ua\da}>0$ (attraction $g_{\ua\da}<0$) in the case of parallel (antiparallel) fields 
has an ``AFM'' nature and is expected to produce high entanglement between the components. 
Because of the monogamy of entanglement \cite{Coffman00}, the entanglement formation between the components 
leads to the destruction of entanglement in each component. 
Thus, the doubled QH states are less stable for such an interaction. 
As shown in Fig.~\ref{fig:phase_ppot}, an intercomponent repulsion $g_{\ua\da}>0$ in the case of (a) parallel fields 
favors all the two-body states with $J\ne 2S$ equally, and thus has a large flexibility 
in the way of forming entanglement between the components. 
This qualitatively explains why a rich variety of spin-singlet QH states emerge for $g_{\ua\da}\approx g$. 
Meanwhile, an intercomponent attraction $g_{\ua\da}<0$ in the case of (b) antiparallel fields 
favors two-body states with small $J$ rather selectively, and is likely to lead to simpler physics.  
In particular, when $g_{\ua\da}=-g<0$, 
the GS is given exactly by the singlet-pairing state $(A_{00}^{\ua\da\dagger})^{N/2}\ket{\mathrm{vac}}$ for all even $N$ \cite{FurukawaUeda14}. 
We note that the formation of larger entanglement for AFM intercomponent couplings than for FM ones is also found 
in the quantum GSs of a binary mixture of spinor Bose-Einstein condensates \cite{Xu12}. 
Generalization of the present argument to other geometries such as a disc and a torus remains as an important open problem. 

%%%%%%%%%%%%%%%%%%%%%%%%%%%%%%%%%%%%%%%%%%%%%%%%%
\section{Summary and outlook}\label{sec:summary}
%%%%%%%%%%%%%%%%%%%%%%%%%%%%%%%%%%%%%%%%%%%%%%%%%

% [ Summary ]--------------------
In this paper, we have determined the QH phase diagram of two-component Bose gases 
in a synthetic magnetic field as shown in Fig.\ \ref{fig:phase}. 
We have revealed a remarkable dependence on the sign of the intercomponent coupling $g_{\ua\da}$: 
while the product states of a pair of QH states are robust for $g_{\ua\da}<0$, 
they are destabilized for moderate $g_{\ua\da}/g$ and a variety of spin-singlet QH states 
with high intercomponent entanglement emerge for $g_{\ua\da}\approx g$. 
We interpret these results in light of pseudopotentials on a sphere. 
The pseudopotential approach also explains recent numerical results 
in two-component Bose gases in antiparallel fields \cite{FurukawaUeda14} 
where a qualitatively opposite dependence on the sign of $g_{\ua\da}$ is found. 

% [ Comments on related systems ]--------------------
It is interesting to ask whether the relationship between the cases of parallel and antiparallel fields 
revealed in the present study and Ref.\ \cite{FurukawaUeda14} applies to more general systems. 
Repellin {\it et al.}\ \cite{Repellin14} have found in lattice models 
that two coupled bosonic Laughlin states with opposite chiralities (i.e., fractional quantum spin Hall states \cite{Bernevig06}) are more robust than 
the ones with the same chiralities for an intercomponent repulsion; 
the case of an intercomponent attraction has yet to be analyzed. 
The stability of fractional quantum spin Hall states against an intercomponent repulsion 
has also been studied in time-reversal-invariant models of spin-$\frac12$ fermions in lattices \cite{Neupert11,LiSheng14} and continuum \cite{ChenYang12}, 
and in a model of strained graphene \cite{Ghaemi12}; 
it is intriguing to compare these systems with their time-reversal-breaking counterparts.  
Further studies in these directions would cross-fertilize two active research fields, 
multicomponent QH systems \cite{MacDonald97} and a strongly correlated regime of spin Hall systems \cite{Bernevig06}. 

\bigskip

% [ Acknowledgments ]--------------------
This work was supported by
KAKENHI Grant Nos. JP25800225 and JP26287088 from the Japan Society for the Promotion of Science, 
a Grant-in-Aid for Scientific Research on Innovative Areas ``Topological Materials Science'' (KAKENHI Grant No. JP15H05855), 
the Photon Frontier Network Program from MEXT of Japan, 
and the Matsuo Foundation.

\appendix

%%%%%%%%%%%%%%%%%%%%%%%%%%%%%%%%%%%%%%%%%%%%%%%%%
\section{Quantum Hall wave functions}\label{app:QH_wvfn}
%%%%%%%%%%%%%%%%%%%%%%%%%%%%%%%%%%%%%%%%%%%%%%%%%
\newcommand{\PLLL}{{\cal P}_\mathrm{LLL}}

Here we summarize QH wave functions discussed in this paper. 

% [ QH states in the scalar case ]--------------------
Let us first review the case of scalar Bose gases. 
We consider a disc geometry, where the LLL orbitals are given by 
$\psi_m(z)\propto z^m\exp[-|z|^2/(4\ell^2)]~(m=0,1,\dots,N_\phi)$ 
with $z=x+iy$ being a complex coordinate. 
In this geometry, a general many-body wave function has a form 
\begin{equation}
 \Psi(\{z_i\})=\Psit(\{z_i\}) e^{-\sum_j |z_j|^2/(4\ell^2)},
 %\exp \left( - \frac{1}{4\ell^2} \sum_j |z_j|^2 \right), 
\end{equation}
where $\Psit(\{z_i\})$ is a symmetric polynomial of the coordinates $\{z_j\}$ of $N$ bosons. 
In the following, we use either $\Psi$ or $\Psit$ to represent each QH wave function. 

The Laughlin wave function \cite{Laughlin83} at the filling factor $\nu=1/2$ is given by 
\begin{equation}\label{eq:Psit_Lau}
 \Psit^\mathrm{Laughlin} = \prod_{i<j} (z_i-z_j)^2.
\end{equation}
This is an exact zero-energy GS for a repulsive contact interaction 
as the amplitude of this wave function vanishes when any two particles come to the same point \cite{Wilkin98}. 
Using this wave function, one can construct the Read-Rezayi series of states \cite{Read99} at $\nu=k/2~(k=1,2,\dots)$, which has an SU$(2)_k$ symmetry. 
Their wave functions can be represented as \cite{Cappelli01}
\begin{equation}\label{eq:SU2_k}
 \Psit^{\mathrm{SU}(2)_k} = \Scal_\group \prod_\group \Psit^\mathrm{Laughlin}.
\end{equation}
Here, the $N$ bosons are first partitioned into $k$ groups with equal populations. 
For each group, we write a Laughlin factor $\Psit^\mathrm{Laughlin}$, and then such factors are multiplied together. 
Finally, we apply the symmetrization operation $\Scal_\group$ over all different ways of dividing the particles into $k$ groups. 
For $k=1$, Eq.~\eqref{eq:SU2_k} clearly gives the Laughlin wave function \eqref{eq:Psit_Lau}; 
for $k=2$,  Eq.~\eqref{eq:SU2_k} is equivalent to the Moore-Read (``Pfaffian'') wave function \cite{Moore91}
\begin{equation}
 \Psit^{\text{MR}} = \Pf \left(\frac{1}{z_i-z_j}\right) \prod_{i<j} (z_i-z_j).
\end{equation}
The SU$(2)_k$ states with $k\ge 2$ exhibit excitations obeying non-Abelian statistics. 
The wave function \eqref{eq:SU2_k} is a unique zero-energy GS of a Hamiltonian consisting of a $(k+1)$-body interaction 
\begin{equation} \label{eq:H_k}
 H_k = \sum_{i_1<\cdots<i_{k+1}} \delta (z_{i_1}-z_{i_2}) \cdots \delta (z_{i_k}-z_{i_{k+1}}). 
\end{equation}
For scalar bosons interacting via a repulsive contact interaction, 
the SU$(2)_k$ wave functions \eqref{eq:SU2_k} have been found to give good approximations to the GSs for small $k$ \cite{Cooper01,Regnault04,Regnault07}. 
On a sphere, the candidate wave functions can be obtained through the replacement $z_i-z_j\to u_iv_j-v_iu_j$ in the above wave functions; 
since the largest power of $z_1$ in Eq.~\eqref{eq:SU2_k} is given by $N_\phi=2(N/k-1)$, these wave functions have the shift $\delta=2$.
On a torus, the SU$(2)_k$ state exhibits topological GS degeneracy of $k+1$. 

Another important series of QH states are Jain's CF states \cite{Jain89,Regnault04} at $\nu=\frac{p}{p+1}~(p=1,2,\dots)$; 
at these filling factors, binding of a unit flux to each boson leads to 
the integer QH states of CFs at the effective filling factors $\nu^*=p$. 
The corresponding wave functions are given by \cite{Jain89,Jain97,Chang05}
\begin{equation}\label{eq:Psi_CF}
 \Psi_{\frac{p}{p+1}}^\mathrm{CF} (\{z_i\}) = \PLLL J(\{ z_i\}) \Phi_p(\{z_i\}),
\end{equation}
where $J(\{z_i\})=\prod_{i<j} (z_i-z_j)$ is the Jastrow factor, 
$\Phi_p(\{z_i\})$ is the Slater determinant obtained by filling exactly $p$ Landau levels, 
and $\PLLL$ is the projection onto the LLL manifold. 
For $p=1$, this wave function reproduces the Laughlin wave function \eqref{eq:Psit_Lau}. 
For $p=2$ and $3$, the wave function \eqref{eq:Psi_CF} (with a slight modification of the projection for technical convenience)
has been confirmed to give good approximations to the GSs for a two-body contact interaction 
in numerical analyses of finite-size systems \cite{Chang05}. 

% [ QH states in the two-component case ]--------------------
Let us now turn to the case of two-component Bose gases studied in this paper. 
Since the QH states for small $g_{\ua\da}/g$ are simply the products of two QH states in the scalar case, 
we here focus on the spin-singlet QH states appearing for $g_{\ua\da}\approx g$. 
The Halperin $(221)$ wave function \cite{Halperin83} at the total filling factor $\nu=1/3+1/3$ is given by
\begin{equation}\label{eq:221}
 \Psit^{221} = \prod_{i<j} (z_i^\ua-z_j^\ua)^2 \prod_{i<j} (z_i^\da-z_j^\da)^2 \prod_{i,j} (z_i^\ua-z_j^\da).  
\end{equation}
The contact interactions in Eq.~\eqref{eq:Hint} vanish for this wave function,  
and therefore Eq.~\eqref{eq:221} is an exact zero-energy GS for arbitrary $g_{\ua\da}\ge 0$ and $g\ge 0$ \cite{Paredes02}. 
Using this wave function, one can construct a series of non-Abelian spin-singlet states  
at $\nu=k/3+k/3$ with integer $k$ \cite{Ardonne99}, which have an SU$(3)_k$ symmetry  
(more generally, the SU$(n+1)_k$ states at $\nu=nk/(n+1)$ can be constructed for $n$-component Bose gases \cite{Reijnders02}). 
On a disc, their wave functions are written as 
\begin{equation}\label{eq:SU3_k}
 \Psit^{SU(3)_k} = \Scal_\group \prod_\group \Psit^{221}.
\end{equation}
Here, as in the Read-Rezayi wave functions \eqref{eq:SU2_k}, 
$N$ bosons are first partitioned into $k$ groups (each with $N/(2k)$ particles in each spin state $\ua,\da$), 
a Halperin $\Psit^{221}$ factor is constructed in each group, 
and then the symmetrization $\Scal_\group$ is carried out. 
The SU$(3)_k$ states with $k\ge 2$ exhibit excitations obeying non-Abelian statistics. 
The wave function \eqref{eq:SU3_k} is again a unique zero-energy GS for a $(k+1)$-body interaction \eqref{eq:H_k} for two components on a disc. 
Since the largest power of $z_1^\ua$ in Eq.~\eqref{eq:SU3_k} is given by $N_\phi=2\left( \frac{N}{2k}-1 \right)+\frac{N}{2k}$, 
these wave functions have the shift $\delta=2$ on a sphere. 
On a torus, the SU$(3)_k$ state exhibits topological GS degeneracy of $(k+1)(k+2)/2$. 
For two-body contact interactions \eqref{eq:Hint} with $g_{\ua\da}\approx g$, 
an indication of $6$-fold GS degeneracy corresponding to the SU$(2)_2$ state
has been obtained numerically for small numbers of particles \cite{Grass12,FurukawaUeda12};  
however, the SU$(2)_2$ state competes with a CFSS state explained below, 
and a recent large-scale simulation based on the iDMRG has provided pieces of evidence 
that the CFSS state is stabilized in the thermodynamic limit \cite{Geraedts17}. 

A series of CFSS states can be introduced at $\nu=\frac{p}{2p\pm 1}+\frac{p}{2p\pm 1}~(p=1,2,\dots)$ \cite{Wu93,Wu13}; 
here, binding of a unit flux with each boson leads to the integer QH states of CFs at $\nu^*=\pm (p+p)$. 
The corresponding wave functions are given by 
\begin{equation}\label{eq:Psi_CFSS}
 \Psi_{\frac{2p}{2p\pm 1}}^\mathrm{CFSS} (\{z_i\}) = \PLLL J(\{ z_i\}) \Phi_{\pm p}(\{z_i^\ua\}) \Phi_{\pm p}(\{z_i^\da\}),
\end{equation}
where $J(\{z_i\})$ is the Jastrow factor for all the particles, 
and $\Phi_{-p}(\{z_i^\alpha \})=\Phi_{p}^*(\{z_i^\alpha \})$. 
For $\nu=2/(2+1)=2/3$, this wave function reproduces the Halperin $(221)$ wave function. 
For $\nu=2/(2-1)=2$, the wave function \eqref{eq:Psi_CFSS} gives the BIQH wave function \cite{Senthil13}, 
which is a good approximation to the GS for two-body contact interactions \eqref{eq:Hint} with $g_{\ua\da}=g$ \cite{Wu13}. 
The BIQH state is particularly intriguing as it is a symmetry-protected topological state of bosons in two dimensions \cite{Chen12,Lu12} 
and exhibits counter-propagating charge and spin modes at the edge \cite{Senthil13}, as numerically demonstrated in Refs.\ \cite{FurukawaUeda13,Wu13}. 
Pieces of evidence for the appearance of the $\nu=4/(4-1)=4/3$ CFSS state in the thermodynamic limit have been obtained 
through the calculations of the shift and the entanglement spectrum in a recent iDMRG simulation \cite{Geraedts17}. 
An indication of the $\nu=4/(4+1)=4/5$ CFSS states has also been found \cite{Wu13}. %(see purple double dotted lines in Fig.\ \ref{fig:gsL}). 
%However, the CFSS state at $\nu=4/3$ competes with the SU$(3)_2$ state, and this competition has been analyzed in Refs~\cite{FurukawaUeda13,Wu13}. 

Indications of gapped states at $\nu=4$ and $6$ have been found in Ref.~\cite{FurukawaUeda13} (see blue dotted lines in Fig.\ \ref{fig:gsL}). 
While we have not achieved appropriate characterizations of these states, 
the real-space entanglement spectrum of the $\nu=4$ state reveals a counterpropagating nature of edge modes, 
suggesting similarities to the BIQH state at $\nu=2$. 
Candidate wave functions for this series of states may be obtained 
by applying a ``grouping and symmetrizing'' procedure as in Eq.~\eqref{eq:SU3_k} to the BIQH wave function; 
however, the relevance of such wave functions to the present system has yet to be clarified. 

%%%%%%%%%%%%%%%%%%%%%%%%%%%%%%%%%%%%%%%%%%%%%%%%%
\section{Pseudopotentials for antiparallel fields}\label{app:pseudopot}
%%%%%%%%%%%%%%%%%%%%%%%%%%%%%%%%%%%%%%%%%%%%%%%%%

\newcommand{\Ncal}{{\cal N}}
\newcommand{\Mcal}{{\cal M}}
\newcommand{\ub}{\bar{u}}
\newcommand{\vb}{\bar{v}}

% [ LLL orbitals ]--------------------
Here we describe the calculation of pseudopotentials 
for pseudospin-$\frac12$ Bose gases in antiparallel magnetic fields on a spherical geometry. 
Such systems have been studied previously \cite{Liu09,Fialko14,FurukawaUeda14}, 
and we basically take the same notations as in Ref.\ \cite{FurukawaUeda14}. 
A related calculation of pseudopotentials for two-species Dirac fermions in antiparallel fields is presented in Ref.\ \cite{Fujita16}. 

We introduce the polar coordinates $(r,\theta,\phi)$ and associated unit vectors $\ev_r,\ev_\theta,\ev_\phi$. 
We place a pseudospin-dependent magnetic monopole of charge $\epsilon_\alpha N_\phi (2\pi \hbar/q)$ with integer $N_\phi\equiv 2S$ at the center of the sphere, 
where $\epsilon_\ua=+1$ and $\epsilon_\da=-1$. 
This monopole produces a magnetic field $\epsilon_\alpha B$ on the sphere of radius $R=\ell \sqrt{S}$. 
For this problem, it is useful to introduce the modified angular momentum 
\begin{equation}\label{eq:ang_mom}
 \Lv_\alpha=\rv \times \left( \pv + \epsilon_\alpha \frac{\hbar S\cot\theta}{r} \ev_\phi \right) -\epsilon_\alpha \hbar S\ev_r, 
\end{equation} 
which obeys the standard algebra of an angular momentum. 
The LLL for a pseudospin-$\alpha$ particle on the sphere corresponds to the subspace in which $\Lv_\alpha$ has the magnitude of $S$. 
The single-particle orbitals in the LLL are given by \cite{Haldane83,Fano86,FurukawaUeda14}
\begin{align}\label{eq:psi_ud}
 \psi_{m\ua} (\rv) = \frac{\bar{v}^{S+m}(-\bar{u})^{S-m}}{\sqrt{4\pi R^2 \Ncal_{S,-m}}},~
 \psi_{m\da} (\rv) = \frac{u^{S+m} v^{S-m}}{\sqrt{4\pi R^2 \Ncal_{Sm}}}, 
\end{align}
for the pseudospin states $\ua$ and $\da$, respectively.  
Here, $m\in\{-S,-S+1,\dots,S\}$ is the eigenvalue of $L_\alpha^z/\hbar$, 
$\rv$ is constrained to the surface of the sphere ($\rv=R\ev_r$), 
and $(u,v)$ and $(\ub,\vb)$ are the spinor coordinates \eqref{eq:uv} and their complex conjugates. 
The normalization factor $\Ncal_{S m}$ is given by
\begin{equation}\label{eq:Ncal_m}
\Ncal_{S m}
=\int \frac{d^2\rv}{4\pi R^2} |u|^{2(S+m)} |v|^{2(S-m)} 
=\frac{(S+m)!(S-m)!}{(2S+1)!}.
\end{equation}
It is worth noting that the orbital $\psi_{m\alpha}$ has the average location 
\begin{equation}\label{eq:psi_cos}
 \bra{\psi_{m\alpha}} \cos\theta \ket{\psi_{m\alpha}} 
 = \int d^2 \rv \cos\theta |\psi_{m\alpha}(\rv)|^2
 = \frac{-\epsilon_\alpha m}{S+1}.
\end{equation}
In particular, the $m=S$ state for $\alpha=\ua$ ($\da$) is localized around the south (north) pole of a sphere. 
This suggests that a pseudospin-$\alpha$ particle in the LLL is, in general, located 
around the direction of $-\epsilon_\alpha \langle \Lv_\alpha \rangle$, as schematically shown in Fig.\ \ref{fig:phase_ppot}(b). 

% [ Pseudopotentials ]--------------------
The pseudopotentials are defined as the eigenenergies of the interaction Hamiltonian \eqref{eq:Hint} for two-body eigenstates. 
Such two-body eigenstates can be calculated through the angular-momentum coupling of Eq.~\eqref{eq:psi_ud} as 
\begin{equation}\label{eq:Phi_ab}
\begin{split}
 \Phi_{JM}^{\alpha\beta} (\rv_1,\rv_2) 
 = \sum_{m_1+m_2=M} &\psi_{m_1\alpha}(\rv_1) \psi_{m_2\beta}(\rv_2) \\
  \times &\bracket{S,m_1;S,m_2}{J,M}, 
\end{split}
\end{equation}
where $\alpha,\beta=\ua,\da$. 
For a general interaction potential $V_{\alpha\beta}(\rv)$, 
the pseudopotentials are given by
\begin{equation}\label{eq:V_J}
 V_J^{\alpha\beta} = \int d^2\rv_1 d^2\rv_2 V_{\alpha\beta}(\rv_1-\rv_2) |\Phi_{JM}^{\alpha\beta}(\rv_1,\rv_2)|^2. 
\end{equation}
Since the right-hand side does not depend on $M$, it is sufficient to consider the case of $M=J$. 
Furthermore, since $V_{\ua\ua}=V_{\da\da}$ and $V_{\ua\da}=V_{\da\ua}$ in the case of our interest, 
we can focus on the cases of $(\alpha,\beta)=(\da,\da)$ and $(\ua,\da)$. 
In these cases, using the expressions of the Clebsch-Gordan coefficients, 
the two-body eigenstates \eqref{eq:Phi_ab} are calculated to be \cite{Haldane83,Fano86,Fujita16}
\begin{subequations}
\begin{align}
&\Phi_{JJ}^{\da\da}(\rv_1,\rv_2)
=\frac{ (u_1v_2-v_1u_2)^{2S-J}(u_1 u_2)^J }{4\pi R^2\Mcal_{SJ}^{1/2}}, \label{eq:Phi_JJ_pp}\\
&\Phi_{JJ}^{\ua\da}(\rv_1,\rv_2)
=\frac{ (\vb_1v_2+\ub_1u_2)^{2S-J} (\vb_1 u_2)^J }{4\pi R^2\Mcal_{SJ}^{1/2}},\label{eq:Phi_JJ_pm}
\end{align}
\end{subequations}
where we introduce the spinor coordinates $(u_i,v_i)$ for $\rv_i~(i=1,2)$ as in Eq.~\eqref{eq:uv}, 
and the normalization factor $\Mcal_{SJ}$ is given by
\begin{equation}
 \Mcal_{SJ} = \Ncal_{J,0} \frac{(2S-J)!(2S+J+1)!}{[(2S+1)!]^2}. 
\end{equation}

% [ Case of contact interactions ]--------------------
We now focus on the case of contact interactions 
$V_{\alpha\beta}(\rv)=g_{\alpha\beta} \delta (\rv)$ with $g_{\ua\ua}=g_{\da\da}=g>0$. 
By substituting Eq.~\eqref{eq:Phi_JJ_pp} into Eq.~\eqref{eq:V_J}, the intracomponent pseudopotential is calculated as
\begin{equation}
V_J^{\da\da}=g\int d^2\rv |\Phi_{JJ}^{\da\da}(\rv,\rv)|^2 =V_{2S}^{\da\da} \delta_{J,2S},
\end{equation}
where
\begin{equation}
V_{2S}^{\da\da}
=\frac{g \Ncal_{2S,2S}}{4\pi R^2\Mcal_{S,2S}}
=\frac{g}{4\pi\ell^2} \frac{(2S+1)^2}{S(4S+1)}. 
\end{equation}
In the limit $S\to\infty$, $V_{2S}$ converges to $g/(4\pi\ell^2)$, 
which coincides with the pseudopotential for zero relative angular momentum 
in a single-component gas on the disk geometry \cite{Cooper08_review}.
Similarly, by substituting Eq.~\eqref{eq:Phi_JJ_pm} into Eq.~\eqref{eq:V_J}, 
the intercomponent pseudopotential is calculated as 
\begin{equation}
V_J^{\ua\da}
=\frac{g_{\ua\da}\Ncal_{J,0}}{4\pi R^2\Mcal_{SJ}}
=\frac{g_{\ua\da}}{4\pi\ell^2} \frac{[(2S+1)!]^2}{S (2S-J)!(2S+J+1)!},
\end{equation}
which gives Eq.~\eqref{eq:Vud_J}. 

%%%%%%%%%%%%%%%%%%%%%%%%%%%%%%%%%%%%%%%%%%%%%%%%%
%References
%%%%%%%%%%%%%%%%%%%%%%%%%%%%%%%%%%%%%%%%%%%%%%%%%


\begin{thebibliography}{99}
\bibitem{Comment_CFSS}
Although the fidelity susceptibility in Fig.\ \ref{fig:fidqh_ovlp_g1}(a) shows a peak around $\gamma=0.21\pi$ for $(N_\phi,N)=(10,12)$, 
this is likely to be due to an accidental competition with the doubled Laughlin states with $(\nu,\delta)=(1,2)$, 
which occurs only for this system size, and is not likely to be related with the transition in the thermodynamic limit . 
For the same reason, we infer that the range of the CFSS state estimated from the data with $(N_\phi,N)=(10,12)$ may need modification at the left boundary. 
Unfortunately, exact diagonalization could not be performed for a larger system.
\end{thebibliography}
\end{document}